\documentclass[aps,prapplied,twocolumn,superscriptaddress]{revtex4-2}

\usepackage{graphicx}
\usepackage{xcolor}
\usepackage{siunitx}

\begin{document}

\title{Nanosecond Photoemission near the Potential Barrier of a Schottky Emitter}

\author{Joshua~L.~Reynolds}
\affiliation{Department of Applied Physics, Stanford University, Stanford, California 94305, USA}
\author{Yonatan~Israel}
\affiliation{Department of Physics, Stanford University, Stanford, California 94305, USA}
\author{Adam~J.~Bowman}
\author{Brannon~B.~Klopfer}
\affiliation{Department of Applied Physics, Stanford University, Stanford, California 94305, USA}
\author{Mark~A.~Kasevich}
\email{kasevich@stanford.edu}
\affiliation{Department of Applied Physics, Stanford University, Stanford, California 94305, USA}
\affiliation{Department of Physics, Stanford University, Stanford, California 94305, USA}

\date{\today}

\begin{abstract}
Nanosecond electron pulses are appealing for ultrafast imaging and electron-gating applications, where tunable currents and narrow energy spreads are desirable. Here, we demonstrate photoemission from a Schottky emitter triggered by nanosecond laser pulses. Using photon energies optimally tuned to the emission potential barrier, we generate pulses containing over $10^5$ electrons with energy spreads below \SI{1}{\electronvolt} with a prompt, single-photon photoemission process. These results are consistent with a theoretical model of laser-triggered electron emission and energetic broadening during propagation and can be widely implemented. 
\end{abstract}

\maketitle

\section{Introduction}
Time-resolved electron microscopy has attracted considerable interest for the study of ultrafast molecular, surface, and bulk dynamics at spatial resolutions below the optical diffraction limit \cite{King2005,Zewail2010,Carbone2011,Sciaini2011,Browning2012,Plemmons2015,Adhikari2017,Liao2017}. To achieve optimal imaging conditions, precise control over the emission and propagation of free electrons is required, and these controls are now also enabling advances in electron-matter interaction experiments \cite{Feist2015,Jones2016,Krueger2018,Kozak2018a,Massuda2018,Wang2020} and microscope designs \cite{Barwick2009,Kruit2016,Schwartz2019,Okamoto2022}. For any electron microscope, the choice of an electron emitter and emission mechanism constrains the achievable imaging conditions due to trade-offs between stability, coherence, and spatial, temporal, and spectral resolutions. 

Short pulses containing large numbers of electrons can be used to decrease exposure times in microscopy and are necessary to generate single-shot images of irreversible dynamics, which require up to $10^9$ electrons per pulse, but Coulomb interactions broaden the spatial and energetic profiles of high-current pulses, increasing aberrations and lowering resolution \cite{Browning2012}. These effects are suppressed in longer pulses, and large numbers of electrons can propagate within nanosecond pulse envelopes while still maintaining the temporal resolution necessary to study processes including phase changes, reaction kinetics, and protein folding \cite{LaGrange2008, Lin2012, Picher2018, Sinha2019}. Furthermore, nanosecond pulses are well-suited for instruments that rely on the fast gating of electrons, such as multi-pass transmission electron microscopes \cite{Juffmann2017, Koppell2019, Klopfer2021}. 

These pulses can be generated by filtering an electron beam in time with a beam blanker, or emission can be triggered by a short laser pulse \cite{Zhang2019}. Blankers are generally integrated with continuous electron sources and can blur or displace electron beams \cite{Zhang2020}. Alternatively, laser-triggering requires optical access to the electron source but introduces different degrees of freedom for controlling the current, temporal duration, and energy spread of the photoemitted pulses.

In this paper, we study electron emission from a Schottky emitter illuminated by nanosecond laser pulses. Optimized to emit bright and stable continuous electron beams \cite{Fransen1999}, these emitters have been laser-triggered with femtosecond pulses in single- and multiphoton photoemission regimes \cite{Yang2010,Bormann2015,Feist2017,Kozak2018,Meuret2019,Olshin2020}. On nanosecond and longer timescales, laser-induced heating of the electron gas has been used to trigger and control emission but can place significant thermal stress on the emitter \cite{Bongiovanni2020} or be restricted to low emission currents \cite{Israel2020}. A single-photon emission regime that does not rely on heating has instead been reached by using nanosecond UV pulses, resulting in broad electron-energy distributions \cite{Olshin2020}. 

In femtosecond photoemission, narrow electron-energy spreads are achieved by matching the photon energy to the emission potential barrier of the source, and longitudinal energy spreads below \SI{500}{\milli\electronvolt} have been demonstrated \cite{Aidelsburger2010,Ehberger2015,Li2017,Karkare2020}. Our work focuses on single-photon photoemission near the potential barrier of a Schottky emitter on nanosecond timescales. We produce pulses containing up to $8 \times 10^5$ electrons with energy spreads below \SI{1}{\electronvolt}, with no evidence of laser-induced thermal effects. Our results are consistent with a theoretical model for laser-illuminated Schottky emitters, which suggests that we are operating with an optimal photon energy. Because these emitters are widely used in custom and commercial microscopes, we anticipate that this regime of nanosecond photoemission can be widely implemented.

\section{Methods}
\subsection{Experimental setup}
\begin{figure}[t]
    \centering
    \includegraphics{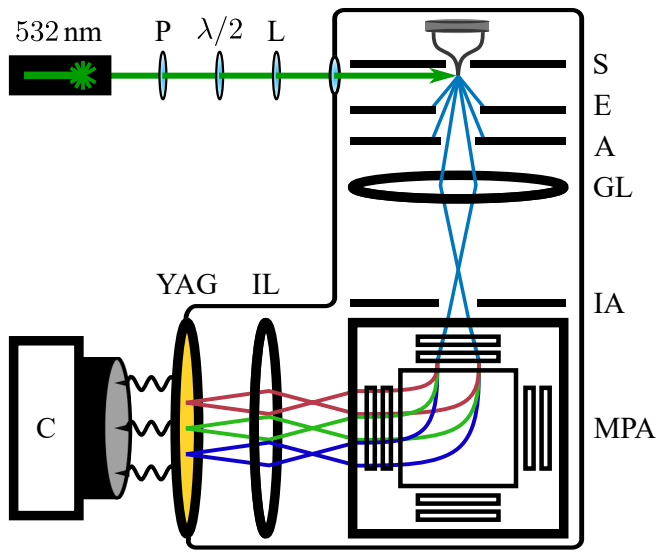}
    \caption{A schematic of the experimental setup. The emission from a ZrO$_\text{x}$/W Schottky emitter at \SI{1300}{\kelvin} is laser-triggered with \SI{0.6}{\nano\second} FW50 pulses from a \SI{532}{\nano\meter} frequency-doubled ND:YAG laser focused onto the tip by a \SI{100}{\milli\meter} lens (L). The laser polarization is controlled with a polarizer (P) and a half-wave plate ($\lambda/2$). The electron optics consist of a suppressor (S), an extractor (E), an anode (A), a gun lens (GL), and an insertable aperture (IA), not drawn to scale. The dispersion induced by deflection through a magnetic prism array (MPA) allows the energy distribution of the electron beam to be relayed by an imaging lens (IL) onto a scintillating YAG:Ce screen (YAG, i.e. yttrium-aluminum-garnet), and the emitted radiation is imaged with a camera (C).}
    \label{fig:Setup}
\end{figure}
Our experiments are conducted with an electron gun (Delong Instruments DIGUN) and magnetic-prism-array system (Electron Optica Monochromator) \cite{Mankos2016}, shown schematically in Fig. \ref{fig:Setup}. The electron source is a \SI{540}{\nano\meter} ZrO$_\text{x}$/W Schottky emitter (DENKA TFE 174), which extends \SI{25}{\micro\meter} past the suppressor electrode. The emitter is heated upon startup to \SI{1820}{\kelvin} to enable the formation of the ZrO$_\text{x}$ layer on the front facet of (100)-oriented tungsten. For photoemission experiments, continuous emission is turned off by cooling the tip to \SI{1300}{\kelvin} over a period of 45 minutes.

We trigger photoemission with \SI{0.6}{\nano\second} FW50, \SI{532}{\nano\meter} (photon energy $E_\text{ph}$ = \SI{2.33}{\electronvolt}) pulses from a frequency-doubled Nd:YAG laser (STANDA-Q-10-SH) operating at \SI{5}{\kilo\hertz}, where FW50 is the temporal duration containing half of the total pulse energy. The incident laser pulses propagate normal to the emitter axis and are linearly polarized, with the polarization direction aligned along the emitter axis unless otherwise noted. A \SI{100}{\milli\meter} focal length lens focuses the laser at the front facet of the emitter to maximize the emission current. Under these operating conditions, the peak intensity of the laser pulse is approximately two orders of magnitude below that used for two-photon photoemission from Schottky emitters \cite{Bormann2015}.

The suppressor and extractor electrodes are separated by \SI{75}{\micro\meter} and are biased by \SI{1.00}{\kilo\volt} and \SI{4.11}{\kilo\volt}, respectively, relative to the emitter. These voltages determine the local field strength at the emitter facet and are held constant to prevent reshaping of the facet surface over time. The final beam energy is set by the bias of the anode, which is located \SI{5}{\milli\meter} beyond the extractor, and is \SI{6}{\kilo\electronvolt} in our experiments. 

Current losses of up to 90\% occur at the extractor, with additional losses occurring at the anode. An insertable aperture located \SI{80}{\milli\meter} below the tip is set to a diameter of \SI{50}{\micro\meter} for all measurements and subtends an effective solid angle of \SI{2e-6}{\steradian}. Approximately 1 in $10^5$ electrons are transmitted through these three optical elements.

After being accelerated, the electrons are focused by the gun lens onto the dispersion plane of the \SI{100}{\milli\meter} magnetic prism array, which is composed of a large magnetic prism surrounded by an array of smaller magnetic coils. The array allows for the correction of astigmatism and distortions in the transmitted beam that would otherwise be present for a single prism \cite{Mankos2016}. The focal spot is \SI{105}{\milli\meter} above the prism array center and is imaged with uniform magnification through the array---with the electron trajectories now bent by \ang{90}---and convolved with the energy distribution of the electrons according to the dispersion of the prism. The energy resolution of this system has been shown to be approximately \SI{10}{\milli\electronvolt} \cite{Mankos2016}. An electrostatic lens relays and magnifies the convolved image onto a scintillating YAG:Ce screen. At \SI{6}{\kilo\electronvolt}, each electron emits approximately 200 photons, and the optical emission is recorded with a camera. 

\subsection{Model of photoemission from a Schottky source}

\begin{figure}[b]
    \centering
    \includegraphics{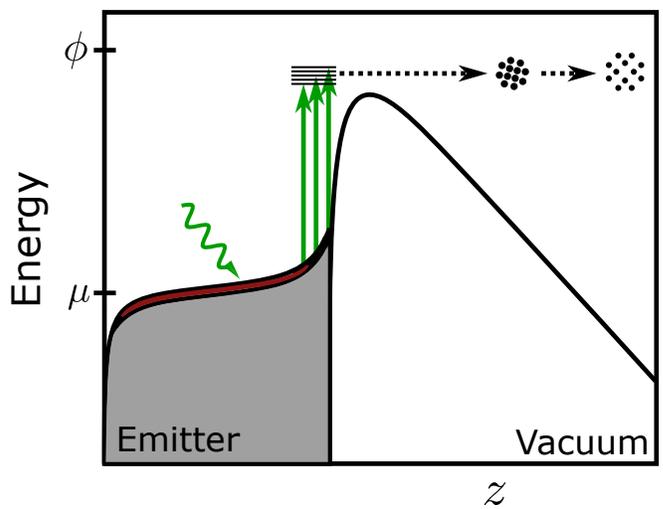}
    \caption{A schematic of the mechanism for near-the-barrier photoemission in one dimension. Most of the emitted current is produced by a single-photon photoemission process enabled by the bent potential barrier in vacuum. Laser-induced heating can increase the population of electrons at higher energies in the tungsten emitter, as shown in red. By matching the photon energy to the barrier height, electron pulses are emitted with narrow intrinsic energy distributions; however, Coulomb interactions after emission still broaden the energy distribution in the longitudinal direction (Boersch effect) and the spatial profile.}
    \label{fig:OverTheBarrier}
\end{figure}

We will now review the physical model for electron emission through potential barriers \cite{Fowler1931, DuBridge1933,Shimoyama1984,Jensen2007}, used here to describe our measurements. As shown in Fig. 2, the one-dimensional potential barrier to emission into the vacuum $U(z)$ for electrons in a Schottky emitter, measured from a zero-energy point at the lowest energy level in the conduction band of the tungsten source, is given by
\begin{equation}
   U(z) = \mu + \phi - \frac{e^2}{16 \pi \epsilon_0 z} - F z e,
\end{equation}
where $\mu$ is the chemical potential, $\phi$ is the work function of the source, $e$ is the unit positive charge, $\epsilon_0$ is the permittivity of free space, and $F$ is the electric field at the metal surface.

Qualitatively, emission occurs when an electron in the metal obtains sufficient energy to tunnel through or overcome the potential barrier, where the electron population at a given energy level is given by Fermi-Dirac statistics, as shown inside the emitter in Fig. \ref{fig:OverTheBarrier}. When illuminated by a laser, photons can be absorbed by the electron gas, increasing the population of electrons with sufficient energy to escape through or over the potential barrier. Furthermore, we highlight that for the nanosecond pulses under consideration, heating of the electron gas to a higher equilibrium temperature is possible, increasing the population at higher energy states, as shown in red. 

One way of differentiating emission regimes is to consider the difference between the photon energy and a barrier height quantified by the effective work function $\phi_\text{eff}$ that results from the lowering of the potential barrier by the applied field $F$, 
\begin{equation}
    \phi_{\text{eff}} = \phi - \sqrt{\frac{e^3 F}{4\pi \epsilon_0}}. 
\end{equation}
Photon energies significantly below $\phi_\text{eff}$ generate negligible currents in the absence of multiphoton processes, and photon energies significantly above $\phi_\text{eff}$ lead to the emission of large currents with broad energy distributions through single-photon absorption. When the photon energy and $\phi_\text{eff}$ are approximately matched in a near-the-barrier regime, single-photon emission still dominates, and the energy spreads of the pulses narrow, yielding ideal operating conditions for imaging. 

In our work, the electric field at the Schottky facet is approximately \SI[per-mode = symbol]{1.1}{\volt\per\nano\meter}. Using measurements of the continuous current emitted by the Schottky source at \SIrange{1800}{1820}{\kelvin}, we measure $\phi$ = \SI{3.1}{\electronvolt}, so $\phi_\text{eff}$ $=$ \SI{1.8}{\electronvolt}, suggesting that near-the-barrier single-photon emission ($\phi_{\text{eff}}\lesssim E_\text{ph}$), as shown in Fig. \ref{fig:OverTheBarrier}, should dominate in our experiments in the absence of continuous emission.

A quantitative description of laser-driven emission from a Schottky source---a photoassisted Schottky tip (PHAST)---based on this model of emission has been proposed by Cook \textit{et al}. \cite{Cook2009}. The PHAST model is applicable to all regimes of Schottky operation, including those that mix continuous thermal-field emission with laser-triggered emission. In the following we restrict our discussion to the regime of laser-triggered emission in which the continuous thermal field emission current from the source is negligible. The current density $J$ emitted in the presence of a cw laser is found by integrating over the emission efficiency for electrons of all kinetic energies normal to the facet surface $E_n$, 
\begin{equation}
    J = \kappa \int_0^{\infty} S(E_n)F(E_n)D'(E_n + E_\text{ph}) \ dE_n,
\end{equation} 
where $S(E_n)$ represents a reduction in the emission current due to scattering in the metal, $F(E_n)$ quantifies the number of electrons with energy $E_n$, and $D'(E_n + E_\text{ph})$ describes the probability of transmission through the potential barrier after the absorption of a photon. The scaling coefficient $\kappa$ quantifies the strength of the illumination and the efficiency of absorption. Importantly, $\kappa \propto I$, the intensity of the illuminating laser. We calculate the current density numerically for a range of experimental conditions, and the full description of the simulation parameters is given in Appendix \ref{SimPar}. The number of electrons per pulse is calculated by dividing the continuous photocurrent by the laser repetition rate, and the intrinsic longitudinal energy spread of the photoemitted electrons is measured as the energy range containing 50\% of the electrons (FW50) for a distribution of electron energies $dJ/dE$. 

\begin{figure}[h]
    \centering
    \includegraphics{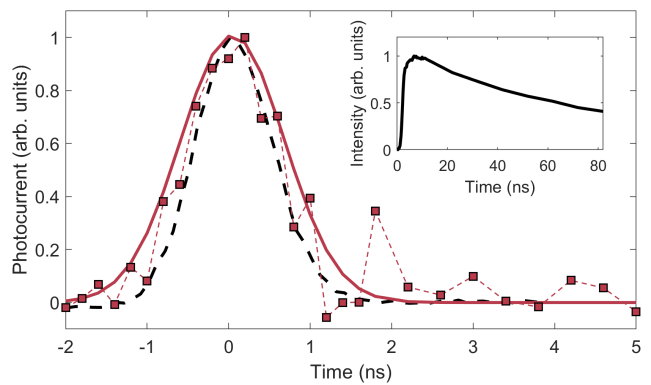}
    \caption{The measured temporal profile of electron pulses (red squares) fit with a $0.86 \pm 0.06$ \si{\nano\second} FW50 Gaussian profile (solid red line). The measured electron and laser (dashed black line) pulse profiles include contributions from the duration of the ICCD gate pulse and the jitter of the electronics. Inset: Temporal response of the YAG:Ce screen, showing a lifetime of $82 \pm 7$ \si{\nano\second}.}
    \label{fig:prompt}
\end{figure}

We note that although previous laser-triggering experiments with nanosecond and microsecond laser pulses have observed heating of the electron gas, the PHAST model assumes a constant emitter temperature. The excited electron gas reaches a quasi-static thermal equilibrium with the emitter lattice on picosecond timescales \cite{Kealhofer2012}, and small temperature increases are possible before the deposited energy is dissipated, resulting in a nonlinear increase in the photocurrent, seen for example in Ref. \cite{Israel2020}, and a broadening of the intrinsic energy distributions of the electrons. These effects are most relevant in low-current and under-the-barrier operating conditions and are neglected in our study of near-the-barrier emission. For longer laser pulse durations, conductive cooling is outpaced by the laser-induced heating, causing larger temperature increases with microsecond rise times \cite{Bongiovanni2020}. Independently, the work function, emission angle, and emission area of a Schottky emitter can also depend on the emitter temperature \cite{Fransen1999, Bronsgeest2013}, and we neglect these effects in our simulations.

As shown schematically in Fig. \ref{fig:OverTheBarrier}, the spatial and energetic profiles of an electron pulse broaden during propagation. The broadening of the intrinsic longitudinal energy spread of an electron pulse by Coulomb interactions during propagation---the Boersch effect \cite{Boersch1954}---is significant at large currents and has been studied for a range of emission conditions \cite{Siwick2002,Passlack2006,Kuwahara2016,Cook2016,Bach2019}. Most of this broadening will occur where the current is high and the electron velocities are low, maximizing electron interaction times. Both of these conditions are satisfied near the emitter immediately after photoemission. Broadening after the extractor hole is negligible due to the large loss of current and high electron velocities.

To estimate the strength of the Boersch effect, we utilize the equations derived for continuous emission from a Schottky source, which include contributions from all four regimes of broadening \cite{Kruit1997}. This approximation of a quasicontinuous beam is justified because the propagation time from the tip to the extractor is just tens of picoseconds, much less than the pulse duration. For a \SI{1}{\nano\second} pulse of known current, we step through the pulse in \SI{50}{\pico\second} intervals and calculate the energetic broadening using the average current in each interval. Assuming that the intrinsic energy distribution is independent of time, the overall energy distribution can then be reconstructed by summing the energy profiles of all such \SI{50}{\pico\second} intervals. In our setup, a \SI{100}{\micro\ampere} continuous beam, corresponding to $3 \times 10^4$ electrons in \SI{50}{\pico\second}, would be broadened by \SI{1}{\electronvolt} during propagation to the extractor. 

\section{Results and Discussion}\label{Sect:results}

Prompt laser-triggered electron emission was verified by imaging the YAG:Ce screen with an ICCD camera with subnanosecond gating capabilities (LaVision Picostar HR). The YAG:Ce exhibits fast rising optical emission ($\lesssim$\SI{1}{\nano\second}) and a long 1/$e$ emission lifetime ($82 \pm 7$ \si{\nano\second}) and thus acts as an integrator of the electron pulse, as shown in the inset of Fig. \ref{fig:prompt}. The $0.86 \pm 0.06$ \si{\nano\second} FW50 pulse profile shown in red in Fig. \ref{fig:prompt} is calculated as the first time derivative of the measured optical intensity as a function of the delay between the ICCD gate pulse and the laser pulse, shown as a black dashed line. Complete details of the measurement are given in Appendix \ref{TimeNotes}. 

The measured electron-pulse duration is comparable to the duration of the laser pulse, confirming that emission results from photon absorption and prompt emission, rather than from longer-lived thermal processes. Temporal broadening due to the longitudinal energy spread of the electrons is negligible, and the longer pulse duration relative to the laser pulse is attributed to the rise time of the YAG:Ce screen. 

\begin{figure}
    \centering
    \includegraphics{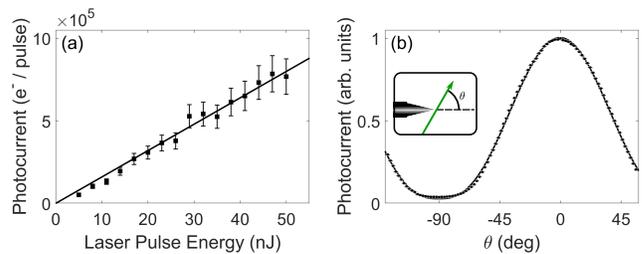}
    \caption{The dependence of the photocurrent on the laser-pulse energy and polarization direction. (a) The measured (squares) and predicted (solid line) photocurrent increases linearly with increasing laser-pulse energy. (b) The relative measured photocurrent (squares) as a function of the angle $\theta$ between the laser polarization (green) and the emitter axis, as shown in the inset, is fitted by a function cos$^{2n}$($\theta$) with $n$ = 1.4 (solid line).}
    \label{fig:polIntScan}
\end{figure}

The dependence of the pulse current on the laser pulse energy and polarization was measured on a CMOS camera (Teledyne FLIR BFS-U3-32S4M-C) and is presented in Fig. \ref{fig:polIntScan}. The continuous current emitted from the Schottky source at high temperatures is independently measured in our instrument and can be used to calibrate the photoemission current. We find that as many as $8 \times 10^5$ electrons per laser pulse are emitted from the tip surface by \SI{50}{\nano\joule} laser pulses (Fig. \ref{fig:polIntScan}a). The measured photocurrent (squares) grows linearly with the laser pulse energy, with error bars indicating statistical uncertainties due to the standard error in the calibration of the absolute emission current and shot noise in the detected current, which is approximately five orders of magnitude smaller than the photocurrent at the emitter facet. This linear dependence of the photocurrent on the laser pulse intensity is consistent with the scaling predicted by the PHAST model and is well matched by the quantitative results of our simulations (solid line).

Importantly, the total photocurrent would not be delivered to the sample plane in a microscope incorporating our electron gun: as noted above, most electrons are lost at the extractor hole, anode, and aperture during propagation. The high gain of the ICCD camera can be used for single-electron counting directly at the YAG:Ce screen, quantifying the number of electrons deliverable to the sample plane. For laser-pulse energies of a few \si{\nano\joule}, pulses containing on order one electron are detected; at lase- pulse energies above \SI{10}{\nano\joule}, tens or more electrons are transmitted through the instrument and can be used to image a sample. Assuming a variance in the number of emitted photons of up to 10\% due to probabilistic emission events such as backscattering \cite{Reimer2008} and that we detect approximately one scintillated photon per electron with the ICCD, the uncertainty in the number of electrons is up to 10\% higher than the shot-noise limit. At higher transmission currents, Coulomb interactions can continue to shape the pulse after the extractor, potentially resulting in different pulse characteristics after the magnetic prism array or in a theoretical sample plane. 

The dependence of the photocurrent emitted by \SI{27}{\nano\joule} laser pulses on the angle between the laser polarization and the tip axis is well fitted by a cos$^{2n}$($\theta$) function with $n = 1.4$ (Fig. \ref{fig:polIntScan}b). Since the photon wavelength is comparable to the characteristic length scale of the emitter, we expect that the additional sharpness ($n > 1$) of the dependence of the photocurrent on the polarization angle is due to the lightning rod effect \cite{Martin2001,Yang2010}. Contributions from multiphoton processes should be negligible due to the low laser intensities used here in comparison to those used in picosecond and femtosecond photoemission experiments, and there is no evidence for polarization-dependent heating of the emitter \cite{Kealhofer2012}, which was previously observed in \cite{Israel2020}.

\begin{figure}
    \centering
    \includegraphics{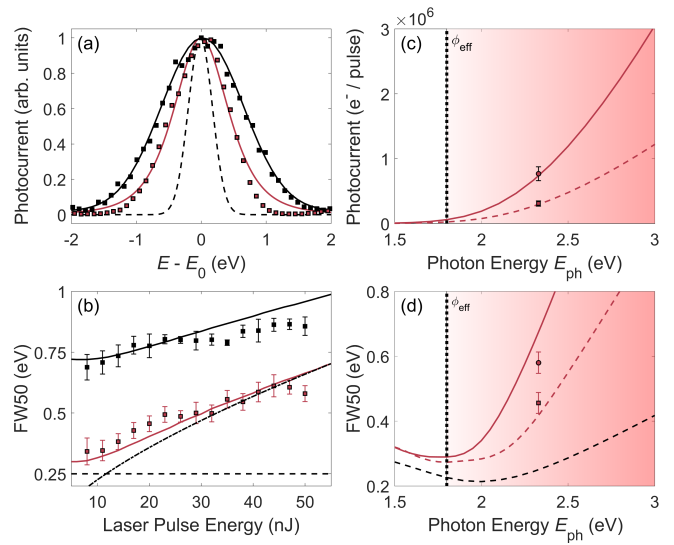}
    \caption{The Boersch effect in nanosecond photoemission. (a) For \SI{50}{\nano\joule} laser pulses, the measured energy profile (black squares) is fitted by a \SI{0.85}{\electronvolt} FW50 Gaussian profile (solid black line). After deconvolving the effective energy width of the focal spot, the energy distribution of the electron bunch (red squares) is recovered from the raw measured spectra. The intrinsic energy distribution predicted by the PHAST model (dashed black line) is broadened by the Boersch effect during propagation, yielding the solid red line. All energies are measured relative to the electron energy at peak instantaneous current $E_0$. (b) The FW50 of the raw (black squares) and deconvolved (red squares) energy profiles increase with increasing laser-pulse energy. The theoretical intrinsic FW50 of the photoemitted pulses (dashed black line) are independent of the laser-pulse energy. The net predicted FW50, with (black line) and without (red line) the prism-induced broadening, increase due to the Boersch effect. At high pulse energies, the FW50 grow consistently with a prediction of Boersch effects in the Holtsmarkian regime (dashed-dotted line). (c),(d) The theoretical predictions and experimental results for (c) the photocurrent and (d) the FW50 (d) are shown for \SI{20}{\nano\joule} (dashed red lines, red squares) and \SI{50}{\nano\joule} (solid red lines, red circles) laser pulses. The dotted vertical line marks the effective work function $\phi_\text{eff}$ in our setup. The color gradient above this energy indicates the onset of the near-the-barrier and then over-the-barrier emission regimes. (d) The intrinsic FW50 (dashed black line) are independent of the laser-pulse energy and are broadened by the Boersch effect at higher currents.}
    \label{fig:energy}    
\end{figure}

The dispersive magnetic prism array enables direct measurements of the energy distributions of the photoemitted pulses. We first measure the transverse positions of the beam on the YAG:Ce screen for a set of known beam energies, where the transverse shift as a function of beam energy calibrates the energy difference as a function of transverse separation. Each two-dimensional pulse profile is then integrated in the direction orthogonal to the dispersion and converted from a spatial distribution to an energetic distribution based on the calibration. The FW50 of this final profile is measured. The predicted intrinsic FW50 of the energy distribution for a photoemitted pulse can be calculated using the PHAST model and is independent of the laser intensity. However, higher laser intensities lead to higher pulse currents, broadening the energy spectrum due to the Boersch effect. 

Example energy profiles are shown in Fig. \ref{fig:energy}(a). The raw experimentally measured energy distribution (black squares) and a Gaussian fit (solid black line) with a FW50 of \SI{0.85}{\electronvolt} are shown for emission triggered with \SI{50}{\nano\joule} laser pulses. The raw measured profile results from the convolution of the spot size of the undispersed beam with the dispersed profile generated by the prism array. While Schottky beams can exhibit non-Gaussian energy distributions \cite{Bronsgeest2007}, we find that the predicted intrinsic and measured total energy profiles are approximately Gaussian, as predicted for large surface field strengths \cite{Bronsgeest2013} and a Gaussian focal spot, respectively.

To remove this broadening due to the finite spot size of the source image, we assume a Gaussian profile for the focused spot at the crossover that has a position and size independent of the photocurrent. We deconvolve a Gaussian with a FW50 of \SI{0.63}{\electronvolt} from the measured spectra to remove the effects of this source image and calculate the FW50 of the recovered energy distributions.

The energy distribution for photoemission triggered by \SI{50}{\nano\joule} laser pulses after deconvolution (red squares) results from the contributions of the intrinsic energy spread and the Boersch effect in the region between the tip and the extractor. After the extractor, Boersch effects, even at beam crossovers, are negligible due to the long pulse durations, high losses, and high kinetic energies. The theoretical intrinsic energy spread (dashed black line) is broadened during propagation from the tip to the extractor (solid red line) for the theoretical current emitted by \SI{50}{\nano\joule} laser pulses. The FW50 of these profiles are \SI{0.25}{\electronvolt} and \SI{0.56}{\electronvolt}, respectively. 

In Fig. \ref{fig:energy}(b), we show that the experimental FW50 of the raw electron-energy distributions (black squares) and the FW50 after deconvolution (red squares) increase with laser pulse energy, or, equivalently, with photocurrent. The error bars indicate the statistical uncertainty $\pm \sigma$ in the estimated FW50 based on Monte Carlo simulations using the measurement noise. The predicted FW50 (solid red line) including contributions from the intrinsic energy spread (dashed black line) and the Boersch effect increase with increasing pulse current. At high currents, the broadening is primarily in the Holtsmarkian regime, and the scaling of the energy width approaches the predicted $I^{2/3}$ scaling (dashed-dotted line). We note that not all electrons experience Boersch effects with strengths in the Holtsmarkian regime, especially as the current decreases, leading to the higher scaling at low laser-pulse energies, where the Boersch effects are closer in magnitude to the FW50 of the intrinsic energy distribution. Despite neglecting variations in the emitter parameters with temperature and aberrations in the electron optics, the model describes the behaviors observed in the experimental data, and we have measured nanosecond pulses with FW50 of as low as $0.34 \pm 0.06$ \si{\electronvolt}. 

Both the intrinsic and broadened energy spreads could be controlled by changing the emission area and propagation dynamics by tuning the suppressor and extractor voltages while maintaining a constant field at the emitter apex, at the risk of reshaping the emitter geometry. Another strategy for further narrowing the energy spread would be to lower $E_\text{ph}$ and move deeper into the near-the-barrier regime, lowering both the intrinsic energy spread and the photocurrent. As shown in Fig. \ref{fig:energy}(c-d), for $E_\text{ph}$ $<$ $\phi_\text{eff}$ (dotted vertical line), photoassisted field emission produces low currents with broad energy spreads. As $E_\text{ph}$ approaches $\phi_\text{eff}$, the intrinsic energy distribution (dashed black line) narrows as a larger number of electrons from a small energy band can be emitted over the barrier. Eventually, higher photon energies again result in broadening energy distributions and large emitted currents. This transition into the near-the-barrier and then over-the-barrier operating regimes is indicated by the red gradients.

The FW50 of the intrinsic energy distribution is not experimentally accessible at high currents due to the Boersch effect, as shown for \SI{20}{\nano\joule} (dashed red lines) and \SI{50}{\nano\joule} (solid red lines) laser pulses in Fig. \ref{fig:energy}(d), where the experimental results at the photon and laser pulse energies reported in this work are given by the red squares and circles. Furthermore, we note that the operating time of laser-triggered Schottky emitters at low temperatures is limited and depends on the photon energy. At 1300~K, ZrO$_\text{x}$ is unable to diffuse to the front facet of the tip, and $\phi$ begins to increase as ZrO$_\text{x}$ is ablated by the laser or contaminated by reactions with residual gas molecules ionized in the vicinity of the tip \cite{Kozak2018}. The timescale of suppression is determined by the energetic separation between the highest energy electrons in the metal and the peak of the Schottky potential barrier relative to the photon energy \cite{Yang2010}. For $E_\text{ph}$ $\gg$ $\phi_\text{eff}$, current losses are minimal for small increases in $\phi$. 

In our setup, the photocurrent is suppressed by more than an order of magnitude after 60 minutes of laser-triggering. This rapid suppression further confirms that the laser photon energy is closely matched to the potential barrier height such that small increases in the work function are sufficient to suppress the observed photocurrent. To compensate for this effect, emitters can be regularly flashed to high temperatures to replenish the ZrO$_\text{x}$ layer and maintain narrow energy spreads \cite{Feist2017, Kozak2018}. 

From this analysis, we conclude that the ideal photon energy sits just above $\phi_\text{eff}$ to enable linear near-the-barrier emission without the need for constant flashing of the emitter. Furthermore, for some applications, higher photocurrent is desirable and can be achieved with the same energy distribution FW50 at slightly higher photon energies. Our experimental results with $E_\text{ph}$ $=$ \SI{2.33}{\electronvolt} satisfy these conditions and suggest the need for accurate characterizations of Schottky source parameters and judicious choices of photon energies when designing instruments that will use nanosecond electron pulses. 

\section{Conclusions}
We characterize a regime of prompt near-the-barrier emission from a Schottky source triggered by nanosecond laser pulses. As many as $8 \times 10^5$ electrons per nanosecond pulse are photoemitted at the emitter facet. These currents exceed those producible from sharper tips \cite{Bormann2010}, and nanosecond laser pulses can be used to reach larger currents than those seen in femtosecond experiments \cite{Yang2010}. Despite these high currents, the extended nanosecond timescale for emission allows narrow electron energy spreads to be maintained, which is important for limiting chromatic aberrations in electron microscopes, and after accounting for losses in our electron gun, we have verified that we can deliver pulses containing single to tens of electrons to future samples.

This single-photon emission regime yields straightforward control over the photocurrent and shows little evidence of thermal nonlinearities or excessive heating of the electron source. Our results are consistent with the PHAST model when energetic broadening due to the Boersch effect is included and suggests that we are operating with an optimal photon energy. These models could prove useful when designing instruments using nanosecond electron pulses. Given the prevalence of Schottky sources, we anticipate that this emission regime will be accessible and valuable for time-resolved imaging, multi-pass electron microscopy, and other fast-gating techniques that manipulate free electrons on nanosecond timescales. 

\begin{acknowledgments}
We would like to thank Marian Mankos of Electron Optica for his help with setting up and optimizing the instrument and for fruitful discussions. This work was conducted as part of the Quantum Electron Microscope collaboration, funded by the Gordon and Betty Moore Foundation. A.J.B. acknowledges support from the Stanford Graduate Fellowship and from the National Science Foundation Graduate Research Fellowship Program under Grant No. 1656518.
\end{acknowledgments}

\appendix

\section{PHAST simulation parameters} \label{SimPar}
Using measurements of the continuous current emitted by the Schottky source at \SIrange{1800}{1820}{\kelvin} and assuming extended Schottky emission, we fit the work function $\phi$ of the source to be approximately \SI{3.1}{\electronvolt} and the emission area to be \SI{1.6}{\micro\meter}$^2$ for a field of \SI[per-mode = symbol]{1.1}{\volt\per\nano\meter} at the emitter facet. The chemical potential $\mu$ is taken to be \SI{10}{\electronvolt}, following \cite{Cook2009}. The reflectivity of the emitter is set to 0.5. The scattering coefficient $S(E_n)$ was found to be approximately constant for the experimental conditions and is set to 1 for all energies. For the laser intensity, we use the average laser power delivered to a focal spot with a \SI{10}{\micro\meter} radius. Calculations are conducted for an emitter operated at \SI{1300}{\kelvin}, and we assume that all parameters are temperature independent. To calculate the energetic broadening due to the Boersch effect, we set the crossover radius equal to the tip radius of \SI{540}{\nano\meter} and take the half-opening angle of emission to be \ang{7}, following \cite{Bronsgeest2013}. The distance from the tip to the extractor is \SI{50}{\micro\meter}. 

\section{Temporal characterization parameters} \label{TimeNotes}
The laser pulse duration at FW50 was measured as $0.64 \pm 0.02$ \si{ns} on a fast photodiode (Hamamatsu G4176). The pulse duration measured on the ICCD camera is $0.65 \pm .04$ \si{\nano\second} and is a convolution of the intrinsic laser pulse duration, the ICCD gating pulse, and the jitter between the pulses, measured to be up to \SI{100}{\pico\second} in our setup. Subtracting the laser pulse duration and jitter in quadrature, the camera gating pulse is measured to be less than \SI{600}{\pico\second}. The electron pulse profile is given by the first time derivative of the optical signal shown in the inset of Fig. \ref{fig:prompt}, where the long emission lifetime of $82 \pm 7$ \si{\nano\second} to 1/$e$ causes the YAG:Ce to act as an integrator of the electron pulse. The jitter and gate duration do not affect the pulse duration when subtracted in quadrature from a Gaussian fit (Fig. \ref{fig:prompt}) that estimates the FW50 pulse duration to be $0.86 \pm 0.06$ \si{\nano\second}.

%

\end{document}